\documentclass[aps,prc,twocolumn,amsmath,floatfix,superscriptaddress]{revtex4}
\usepackage{color,graphicx,ulem,subfig,soul,multirow}
\usepackage{mathptmx}                
\usepackage{dcolumn}                 
\usepackage{bm}                      
\usepackage{epsfig,amsmath}
\usepackage{xcolor}

\newcommand{\vect}[1]{\boldsymbol{#1}}

\begin{document}

\title{Nuclear skin and the curvature of the symmetry energy}

\author{Ad. R. Raduta\thanks{araduta@nipne.ro}}
\affiliation{National Institute for Physics and Nuclear Engineering (IFIN-HH), RO-077125, Bucharest-Magurele, Romania}
\author{F. Gulminelli\thanks{gulminelli@lpccaen.in2p3.fr}}
\affiliation{Universit\'e de Caen Normandie, ENSICAEN, LPC, UMR6534, F-14050 Caen, France}

\date{\today}

\begin{abstract}
The effect of correlations between the slope and the curvature of the symmetry energy on 
ground state nuclear observables is studied within the extended Thomas-Fermi approximation.
We consider different isovector probes of the symmetry energy, with a special focus on the neutron 
skin thickness of $^{208}{\rm Pb}$. 
We use a recently proposed meta-modelling technique to generate a large number of equation of state models, 
where the empirical parameters are independently varied. The results are compared to a set of calculations 
using 17 different Skyrme interactions. We show that the curvature parameter plays a non-negligible role 
on the neutron skin, while the effect is reduced in Skyrme functionals because of 
the correlation with the slope parameter. 
\end{abstract}

\maketitle

\section{Introduction}
The determination of the nuclear matter equation of state (EoS) is an extremely lively issue in 
modern nuclear physics and astrophysics. The biggest uncertainties concern high density and strongly asymmetric matter, 
where the EoS determination is of outermost importance for the understanding of a large variety of 
astrophysical phenomena involving compact stars \cite{oertel16,lattimer12}. Observational measurements of 
neutron star mass and radii start to provide compelling constraints to the behavior of 
high density matter \cite{steiner16,steiner14}, including the very recent multi-messenger observation 
of a neutron star merger \cite{LIGO}, where the EoS has a direct impact on the gravitational wave form 
mainly through the tidal polarizability parameter \cite{steiner15}. In this context, tight constraints 
coming from controlled nuclear experiments are extremely important, particularly concerning the 
isovector part of the EoS, the so-called symmetry energy \cite{epja50,Lattimer16}. 
A huge literature is devoted to the determination of the symmetry energy at saturation ($E_{sym}$) 
and its slope ($L_{sym}$) by comparing selected isovector observables to EoS models issued 
from different energy density functionals (EDF) \cite{tsang_rep,Kortelainen12}.
These studies have convincingly shown that a strong linear correlation exists between the $L_{sym}$ parameter and 
the neutron skin thickness \cite{furnstahl,rocamaza12}. This latter can be measured directly from 
parity violating electron scattering \cite{Abrahamyan_2012} and pion photoproduction \cite{tarbert}, 
or probed via various isovector modes of collective excitations \cite{Trzcinska_PRL87,Tamii_2011,rossi,birkhan,carbone,henshaw}.
A good correlation is also typically observed with the $E_{sym}$ parameter \cite{Nazarewicz14,papazoglou} 
and qualitatively explained by the fact that the behavior of the symmetry energy is, to a first order approximation, 
linear in density in the subsaturation regime \cite{RocaMaza_JPG_2015}. 
However, this correlation is somewhat blurred when different families of mean field models 
are compared \cite{Mondal_PRC93}, showing that some residual model dependence exists. 
The careful study of Ref. \cite{Mondal_PRC93} shows that this difference can be ascribed to different 
nucleon density distributions in the surface region. In turn, this can be due both to different 
surface properties of the functionals, or to different behaviors of the symmetry energy at subsaturation, 
that is to deviations from the linear approximation. 

To progress on this issue, it is important to assess the role of the curvature of the symmetry energy ($K_{sym}$) 
on isovector probes such as the nuclear skin. Little attention was paid to this parameter in the literature 
until recently \cite{Mondal17,Tews17}, mainly due to the fact that it cannot be easily varied within a specific EoS model, 
because the functional form of the EoS imposes a correlation with the low order parameters $E_{sym}$ and $L_{sym}$. 
Still, if $K_{sym}$ is of secondary role for nuclear structure observables, it is the main source of 
uncertainty when extrapolating the laboratory constraints to the high density domain relevant for 
neutron star physics \cite{JM2}.

To perform this study, we use a recently proposed meta-modelling approach to the EoS \cite{JM1}, 
where a large number of different EoS models can be generated without any a-priori correlation among 
the different empirical parameters. Following Ref. \cite{Debi}, ground state observables are calculated 
within the Extended Thomas Fermi (ETF) approximation, with the addition of a gradient term as an 
effective parameter representing the different surface properties of the different models. 

We show that the $K_{sym}$ parameter plays a non-negligible role in the nuclear skin as well as in
the differences of the proton radii of mirror nuclei 
and that the uncertainty on this parameter partially blurs the correlation with the symmetry energy slope.

The paper is organized as follows: the different energy functionals are briefly reviewed in section \ref{sec:formalism}, 
as well as the  ETF approximation used to calculate nuclear observables. 
In Section \ref{sec:results}, after discussing the overall performance of the ETF approximation 
on the Pb isotopic chain, we show  our main results concerning the correlations between the 
different isovector observables and the empirical EoS parameters. 
Finally conclusions are drawn in section \ref{sec:conclusions}. 

\section{Formalism}\label{sec:formalism}

\subsection{Skyrme EDF}

The most extensive calculations of nuclear observables and their correlations with EoS parameters 
have been performed using Skyrme EDF \cite{Dutra_Skyrme}.

The nuclear Skyrme energy density is expressed in terms of local nucleon densities $n_q(\vect{r})$,
kinetic energy densities $\tau_q(\vect{r})$ and spin-orbit densities $\vect{J}_q(\vect{r})$
defined by~\cite{Vautherin_72} 
\begin{eqnarray}
n_q(\vect{r})&=&\sum_{\nu, s}|\phi_{\nu}(\vect{r},s,q)|^2 n_{\nu}^q,\nonumber \\
\tau_q(\vect{r})&=&\sum_{\nu, s}|\nabla \phi_{\nu}(\vect{r},s,q)|^2 n_{\nu}^q,\nonumber \\
\vect{J}_q(\vect{r})&=&(-i)\sum_{\nu, s, s'} \phi_{\nu}^*(\vect{r},s',q)
\nabla \phi_{\nu}(\vect{r},s,q) \times \langle s'|\sigma|s\rangle n_{\nu}^q,
\end{eqnarray}
where $\phi_{\nu}(\vect{r},s,q)$ represent the single-particle wave functions with orbital and
spin numbers $\nu$ and $s$, $q=n,p$ indexes the nucleonic species
and $n_{\nu}^q$ are the occupation numbers.
The functional form of the EDF is generated by a mean-field calculation with 
an effective zero range momentum dependent  pseudo-potential, augmented of a density dependent term.
Standard pseudo-potentials, as the ones considered hereafter, depend on 10 parameters.
The values of these parameters are typically determined by fits of experimental ground-state properties of 
spherical magic and semi-magic nuclei (e.g. binding energy, root mean square (rms) radius of the charge distribution, 
spin-orbit splitting, isotope shifts, surface thickness, breathing mode energy, etc.) 
and/or properties of symmetric nuclear matter (energy $E_{sat}$ and density $n_{sat}$ at saturation, 
compression modulus $K_{sat}$, symmetry energy $E_{sym}$) 
and/or equation of state of pure neutron matter as predicted by ab-initio models.
These parameters vary largely from one Skyrme model to another.  
Properties of nuclear matter (NM) can be expressed analytically in terms
of the same parameters \cite{Dutra_Skyrme}.

In the following, 17 Skyrme EDFs will be employed: 
SKa~\cite{SKab}, SKb~\cite{SKab}, Rs~\cite{Rs}, SkMP~\cite{SkMP}, SLy2~\cite{SLY2and9}, SLy9~\cite{SLY2and9},
SLy4~\cite{SLY4}, SLy230a~\cite{SLY230a}, 
SkI2~\cite{SkI2-5}, SkI3~\cite{SkI2-5}, SkI4~\cite{SkI2-5}, SkI5~\cite{SkI2-5}, SkI6~\cite{SkI6},
SKOp~\cite{SKOp}, SK255~\cite{SK255and272}, SK272~\cite{SK255and272} and KDE0v1~\cite{KDE0v}. 
The extent to which they fulfill various constraints that have been obtained from experiment or microscopic
calculations during the last decade \cite{Lattimer_Lim}
has been thoughtfully investigated in Ref. \cite{Fortin16} 
in the context of unified equations of state for neutron star matter.
Their values of saturation density of symmetric nuclear matter (SNM),
energy per particle and compression modulus of symmetric saturated matter span relatively
narrow ranges  
$0.1512 \leq n_{sat} \leq 0.1646$ ${\rm fm}^{-3}$,
$ -16.33 \leq E_{sat} \leq -15.52 $ MeV,
$222.40 \leq K_{sat} \leq 271.5 $ MeV,
as these quantities are relatively well constrained.
Larger domains are explored by the symmetry energy,
$29.54 \leq E_{sym} \leq 37.4$ MeV, and, especially, its slope and curvature
$44.3 \leq L_{sym} \leq 129.3$ MeV, $-127.2 \leq K_{sym} \leq 159.5$ MeV.

It is worthwhile to notice that the functional form of the Skyrme energy density  leads to  
correlations between the different EoS parameters. Indeed 5 independent parameters govern the density dependence
of the EDF (and 2 additional ones determine the density dependence of the effective masses). 
If the lowest order EoS parameters are fixed, namely $E_{sat}$, $n_{sat}$, $K_{sat}$, $E_{sym}$, $L_{sym}$, 
the higher order parameters can be analytically expressed as a function of those fixed quantities. 
In particular, Skyrme EDFs show a clear correlation between the slope $L_{sym}$ and the curvature 
$K_{sym}$ of the symmetry energy at saturation, which are a-priory independent EoS parameters. 

\begin{figure}
\begin{center}
\includegraphics[angle=0, width=0.99\columnwidth]{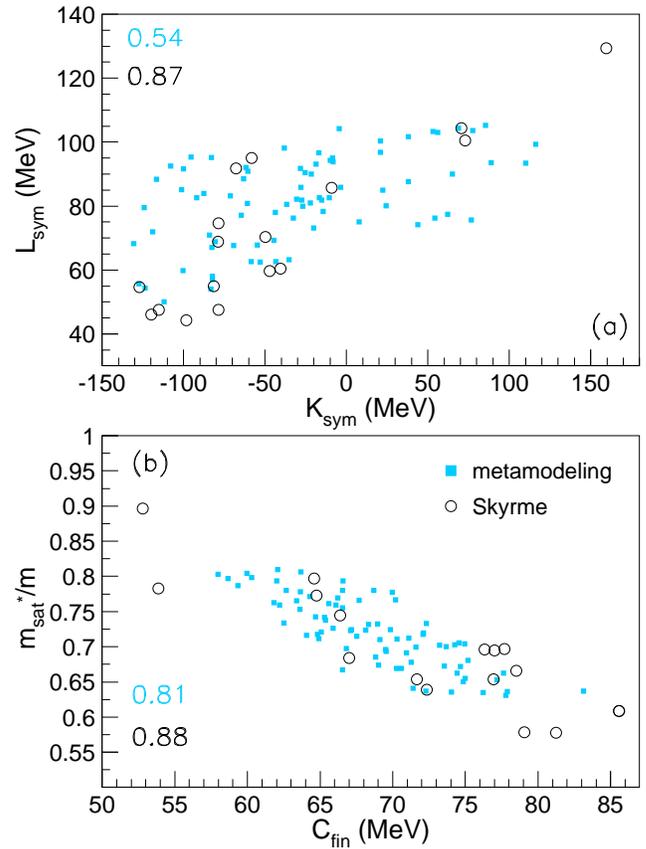}
\end{center}
\caption{Correlations between EoS parameters. 
Numbers on l.h.s. of each plot correspond to Pearson correlation coefficients between the parameters
plotted on the axis. Upper (lower) values: meta-model (Skyrme). 
}
\label{fig:EoSParamCorr}
\end{figure}

This correlation, which obviously affects the extrapolation of the EoS to super-saturation densities, 
is graphically illustrated in the top panel of Fig. \ref{fig:EoSParamCorr} (open circles). 
Its Pearson correlation coefficient\footnote{The Pearson correlation coefficient between two variables $X$ and $Y$ is defined by
$C(X,Y)=\left(\langle XY \rangle -\langle X \rangle \langle Y \rangle\right)/\sigma(X) \sigma(Y)$.}
is $C(K_{sym},L_{sym})=0.87$.
It was recently shown that this correlation is observed in a large class of functionals and might therefore 
be physically founded \cite{Mondal17,Tews17}, even if its origin is not fully understood.  

Another interesting non-trivial correlation is found between 
the effective nucleon mass at saturation, $m_{sat}^*$, 
and the isoscalar-like finite size parameter $C_{fin}$ (see section IIB and Ref. \cite{Debi}).
This correlation is illustrated in the bottom panel of Fig. \ref{fig:EoSParamCorr}.
Its Pearson correlation coefficient is $C(m_{sat}^*,C_{fin})=0.88$. 
As already discussed in ref. \cite{Debi}, this correlation is probably induced by the parameter fitting protocol 
of Skyrme functionals. Indeed  $m_{sat}^*$ and $C_{fin}$ are related to non-local terms in the EDF which have an 
opposite effect on the surface energy, and neither of them  plays a role on the determination of EoS parameters: 
for a given set of EoS parameters a similar overall reproduction of binding energies over the nuclear chart 
can be obtained with compensating effects of the non-local terms.

\subsection{Meta-modelling of the EDF}

A theoretical calculation of a nuclear observable depends, besides the EoS, on the functional form assumed for the EDF 
as well as on the many-body technique employed.
To assess the model dependence due to the functional form of the EDF, one should consider different families of models 
with similar values for the EoS parameters. To this aim, a meta-modelling technique was proposed in Ref. \cite{JM1} 
and extended to finite nuclei EDF in Ref. \cite{Debi}.
Varying the parameters of the meta-modelling, a large number of EoS from different families of mean-field EDF can be generated. 
Moreover, novel density dependencies that do not correspond to existing functionals but do not violate any empirical constraint, 
can be also explored \cite{JM1}. The inclusion of a single gradient term provides a minimal flexible EDF for finite nuclei, 
with performances on nuclear mass and radii comparable to the ones of full Skyrme functionals \cite{Debi}.
The exploration of the meta-modelling parameter space thus allows a full estimation of the possible model dependence 
of the extraction of EoS parameters from nuclear ground state observables, due to the choice of the EDF.

The potential energy per baryon is expressed as a Taylor expansion around saturation of symmetric nuclear matter
in terms of the density parameter  $x=(n - n_{sat})/(3 n_{sat})$, 
\begin{equation}
e_{pot}(x,\delta) = \sum_{\alpha=0}^N (a_{\alpha 0} + a_{\alpha 2} \delta^2) \frac{x^{\alpha}}{\alpha!} u_{\alpha} (x) \,,
\label{eq:epot}
\end{equation} 
where the functions $u_{\alpha}(x)$ represent a low density correction insuring  a vanishing energy in the limit of vanishing density, 
without affecting the derivatives at saturation.

To correctly reproduce with a limited expansion order $N$ existing non-relativistic (Skyrme and ab-initio) 
and relativistic (RMF and RHF) EDFs up to total densities $n=n_n+n_p \approx 0.6$ fm$^{-3}$, 
and  isospin asymmetries $\delta=(n_n-n_p)/n$ ranging from symmetric matter $\delta=0$ to pure neutron matter $\delta=1$,
the functional is supplemented by a kinetic-like term adding the expected $n^{2/3}$ dependence at low densities, 
as well as the contribution of higher orders in the $\delta$ expansion, as:
\begin{equation}
 e_{kin}(x,\delta) = \frac{t_{sat}^{FG}}{2} (1+3 x)^{2/3} \left[ (1+\delta)^{5/3} \frac{m}{m_n^*} + (1-\delta)^{5/3}\frac{m}{m_p^*}  \right]~,
\end{equation}
where $t_{sat}^{FG} = \left( 3 \hbar^2 \right)/\left( 10 m \right) \left(3 \pi^2/2 \right)^{2/3} n_{sat}^{2/3}$
is the energy per nucleon of a free symmetric Fermi gas at nuclear saturation,
$m$ stands for the nucleon mass and $m_q^*$ denote the effective mass of the nucleons $q=n,p$.
For more details, see model ELFc in Ref.~\cite{JM1}.

In the present work, we only consider subsaturation matter and, to avoid proliferation of unconstrained parameters, 
we limit the expansion to $N=2$, which was shown to be enough to get a fair reproduction of nuclear masses \cite{Debi}. 
The possible influence of higher order parameter  is left for future work. 
When only average nuclear properties (e.g. binding energies and rms radii of neutron and
proton distributions) are calculated, isoscalar and isovector finite-size and spin-orbit interactions 
can be fairly well described by a single isoscalar-like density gradient term \cite{Debi} of the form
$C_{fin} \left(\nabla n_n+\nabla n_p \right)^2$.
For the sake of convenience only this isoscalar density gradient will be considered in this work.
Following Ref. \cite{Debi}, we also neglect the effective mass splitting between neutrons and protons.
The meta-modelling parameters are then directly linked to the usual first and second order empirical parameters of the EoS by:

\begin{eqnarray}
a_{00} &=& E_{sat} - t_{sat}^{FG} ( 1 + \kappa_{sat}) \\
a_{10} &=& - t_{sat}^{FG} ( 2 + 5 \kappa_{sat}) \\
a_{20} &=& K_{sat} - 2 t_{sat}^{FG} ( -1 + 5 \kappa_{sat} ) \\
a_{02} &=& E_{sym} - \frac{5}{9} t_{sat}^{FG} ( 1 + \kappa_{sat}) \\
a_{12} &=& L_{sym} - \frac{5}{9} t_{sat}^{FG} ( 2 + 5\kappa_{sat}) \\
a_{22} &=& K_{sym} - \frac{10}{9} t_{sat}^{FG} ( -1 + 5\kappa_{sat}) 
\end{eqnarray}
where $\kappa_{sat} = m/m^*_{sat} - 1$.

\begin{table*}[tb]
\centering
\setlength{\tabcolsep}{2pt}
\renewcommand{\arraystretch}{1.2}
\begin{ruledtabular}
\caption{
Average and standard variation of the different parameters of the phenomenological EDF,
calculated based on 51 Skyrme interactions and 15 relativistic mean-field interactions 
(see table IV in Ref. \cite{JM1}).
}
\label{tab:EDF}
\begin{tabular}{ccccccccc}
  Parameter $\{ P_\alpha \} $ & $ n_{sat}$ & $E_{sat}$ & $K_{sat}$ & $E_{sym}$ & $L_{sym}$ & $K_{sym}$ & $m^*_{sat}/m$ & $C_{fin}$ \\
    {} & (${\rm fm}^{-3}$) & (MeV) & (MeV) & (MeV) & (MeV) & (MeV) & & (${\rm MeV fm}^5$)\\
\hline
Average $\langle \{P_\alpha\} \rangle$ & 0.1543 & -16.03 &  251 & 33.30 & 76.6 & -3 & 0.72 & 75 \\ 
Standard deviation $\sigma_{\alpha}$ & $\pm$0.0054 & $\pm$0.20 & $\pm$29 & $\pm$2.65 & $\pm$29.2 & $\pm$132 & $\pm$0.09 & $\pm$25 \\ 
\end{tabular}
\end{ruledtabular}
\end{table*}

Different EDF models for nuclei are generated by largely and evenly exploring the parameter space
$\{P_\alpha\} = \{ n_{sat}, E_{sat}, K_{sat}, E_{sym}, L_{sym}, K_{sym}, m_{sat}^*, C_{fin}\}$.
For a given model, the ground state nuclear energies and radii are calculated in the 
extended Thomas Fermi approximation at second order, as detailed in the next section.
We retain for the subsequent analysis only the models $\{ P_\alpha \}$ which provide 
a fair description of the experimental binding energies of the spherical magic nuclei:
(40,20), (48,20), (48,28), (58,28), (88, 38), (90, 40), (114, 50), (132, 50), (208, 82)
and charge radii of (40,20), (48,20), (58,28), (88, 38), (90, 40), (114, 50), (132, 50), (208, 82).
The absence of the nucleus (48,28) in the second list is due to the fact that its experimental 
charge radius is not yet available. 
We recall that this set of data represents the core of nuclear properties on which the parameters of
many Skyrme interactions have been fitted. The limitation to spherical nuclei is obviously due
to the simplifying spherical approximation of most approaches, including ours.
Specifically, retained EDFs correspond to sets of parameters $\{ P_\alpha\}$ which provide 
$\chi(B) \leq 5$ MeV and $\chi(R_{\rm ch}) \leq 0.10$ fm.
The mimimum values here obtained for standard deviation of masses and charge radii are 
2.7 MeV and, respectively, $2.07 \cdot 10^{-2}$ fm.
As usual in the literature, the chi-square function is defined as 
$\chi^2(X)=\sum_{i=1}^N \left(X_{{\rm ETF}(i)}-X_{{\rm exp}(i)} \right)^2 /N$.
The accepted values of standard deviation on mass are typically one order of magnitude larger than
the lowest value in the literature, 0.5 MeV, which corresponds to more than 2350 nuclei and has been obtained
in the framework of a Hartree-Fock-Bogoliubov (HFB) mass model \cite{Goriely_2013}.

The variation domain of each parameter is obtained by considering the dispersion of the corresponding values
in a large number of relativistic and non-relativistic mean-field models, see Ref. \cite{JM1}.
The precise frontiers of this domain depend on the number of models considered and their selection criteria, 
and is therefore somewhat arbitrary. However, a variation of the borders of the parameter space might affect 
the overall dispersion in the predictions of the meta-model, but not the quality of the correlations 
among parameters and observables, which is the scope of the present work.
 
The  domain considered for each parameter $P_\alpha$ is reported in Table \ref{tab:EDF} 
in terms of average value and standard deviation.
Good/poor experimental constraints on 
$n_{sat}$, $E_{sat}$ and $E_{sym}$ on one hand and
$K_{sat}$, $L_{sym}$ and $K_{sym}$ on the other hand
lead to narrow/wide variation domains of these variables.

As a first application of the meta-modelling, we can investigate the model dependence of the correlations among 
empirical parameters observed in the previous section for the Skyrme EDFs.

The only significant correlation that was found in the different models generated by the meta-modelling technique 
after application of the mass and radius filter, is the one between $m_{sat}^*$ and $C_{fin}$, 
as shown by solid squares in the bottom panel of Fig. \ref{fig:EoSParamCorr}. 
The value of the correlation coefficient, $C= 0.81$, is close to our previous calculations 
using a simplified version of the extended Thomas Fermi approach \cite{Debi}, 
and also to the correlation coefficient of Skyrme pseudo-potentials. 
This confirms that the Skyrme correlation comes from the physical constraint of mass reproduction, 
and is largely independent of the EDF model.

Conversely, only a poor correlation between $L_{sym}$ and $K_{sym}$ emerges from the meta-modelling after 
application of the mass constraint, see solid squares in the top panel of Fig. \ref{fig:EoSParamCorr}.
This suggests that the origin of that correlation observed in different functionals \cite{Mondal17,Tews17} 
is not due to the constraint of mass reproduction.

\subsection{The Extended Thomas-Fermi approximation with parametrized density profiles}

For a given EDF model, average properties of atomic nuclei can be  reasonably well described within the 
Extended Thomas-Fermi (ETF) approximation \cite{Brack_1985}. 
In this work, we will limit ourselves to the second order expansion in $\hbar$ and 
to parametrized density profiles in spherical symmetry, such as to limit the number of variational parameters. 
Because of these approximations, the degree of reproduction of experimental data 
is not comparable to the one of dedicated fully quantal HFB calculations 
\cite{Goriely_2013}, 
and more realistic calculations will definitely have to be performed in order to determine 
EoS parameters in a fully quantitative way. 
Still, the complete exploration of the parameter space is not affordable with these more sophisticated 
many body techniques, and we believe that an ETF meta-modelling is sufficient to extract the 
correlations between EoS parameters and the neutron skin.

In the ETF framework, the energy of an arbitrary distribution of nucleons with densities
$\{n_n(\vect{r}), n_p(\vect{r})\}$ 
is given by the volume integral of the energy density according to:
\begin{equation}
E_{tot}=\int d \vect{r} \left( e_{nuc}\left[n_n, n_p\right] 
+ e_{Coul} \left[ n_p \right] \right), 
\end{equation}
where the first term stands for the nuclear energy and the second for the electrostatic contribution.

At second-order in the $\hbar$ expansion, the nuclear energy density functional writes
\begin{equation}
e_{nuc}\left[n_n, n_p \right]
=\sum_{q=n,p} \frac{\hbar^2}{2 m_q^*} \tau_{2q} + e_{TF},
\label{eq:enuc}
\end{equation}
where $e_{TF}$ is the Thomas-Fermi approximation of the chosen nuclear EDF model, 
which can depend on local densities $n_q$ as well as on density gradients $\nabla n_q$ 
and currents $\vect{J}_q$
and $\tau_{2q}$ is the (local and non-local) density dependent correction 
arising from the second order $\hbar$ expansion of the kinetic energy density operator. 

The Coulomb energy density is expressed as \cite{Onsi_PRC2008},
\begin{eqnarray}
 e_{Coul} [n_p]=\frac{e^2}{2} n_p(\vect{r}) \int \frac{n_p(\vect{r'})}{|\vect{r}-\vect{r'}|} d \vect{r'}
-\frac{3 e^2}{4} \left (\frac{3}{\pi}\right)^{1/3} n_p^{4/3}(\vect{r}),
\label{eq:exactCoulomb}
\end{eqnarray}
where the Slater approximation has been employed to estimate the exchange Coulomb energy density.

The ground state is determined by energy minimization using parametrized 
neutron and proton distributions.
For a generic nucleus with $N$ neutrons and $Z$ protons 
and under the simplifying approximation of spherical symmetry, 
these are customarily parametrized as Wood-Saxon (WS) density profiles,
\begin{equation}
n_{q}^{WS} (r)=\frac{n_{bulk, q}}{1+\exp\left[ \left(r-R_q^{WS} \right)/a_q \right]},
\label{eq:DensityProfile}
\end{equation}
where $n_{bulk, q}$ is linked to the central density of the 
$q=n,p$ distribution,
and
$R_q^{WS}$ and $a_q$ respectively stand for radius and diffuseness parameters. 
With the extra condition of particle number conservation,
\begin{equation}
Z=4 \pi \int_0^\infty d r~ r^2~n_p(r),~ N=4 \pi \int_0^\infty d r~ r^2~n_n({r})
\label{eq:PartNrConserv}
\end{equation}
 only four variables out of six are independent.
In the variational calculation of the ground state,
we make the choice of variating $\{n_{bulk, q}, a_q ; q=n,p\}$,  while
$R_q^{WS}$ are obtained from Eq. (\ref{eq:PartNrConserv}).

The only experimental observables related to the distribution of matter are the root mean squared (rms)
radius of the charge distribution and, with larger error bars, neutron skin thickness.
Rms radius of the charge distribution is defined as the rms radius of the proton distribution
corrected for the internal charge distribution of the proton $S_p$=0.8 fm,
\begin{equation}
\langle r_{ch}^2\rangle^{1/2}=\left[ \langle r_p^2 \rangle+S_p^2\right]^{1/2}.
\label{eq:rch}
\end{equation}

Neutron skin thickness is defined as the difference in the neutron-proton rms radii,
\begin{equation}
\Delta r_{np}=\langle r_n^2 \rangle^{1/2}-\langle r_p^2 \rangle^{1/2},
\label{eq:skin}
\end{equation}
and, as demonstrated in Ref. \cite{Mondal_PRC93}, it can be decomposed with good accuracy
into a bulk contribution, 
\begin{equation}
\Delta r_{np}^{bulk}=\sqrt{\frac35} \left[ \left( R_n^{WS}- R_p^{WS}\right) +
\frac{\pi^2}{3} \left(\frac{a_n^2}{R_n^{WS}}-\frac{a_p^2}{R_p^{WS}} \right) \right],
\label{eq:skin_bulk}
\end{equation}
and a surface contribution,
\begin{equation}
\Delta r_{np}^{surf}=\sqrt{\frac35} \frac{5 \pi^2}{6} \left( \frac{a_n^2}{R_n^{WS}}-\frac{a_p^2}{R_p^{WS}} \right).
\label{eq:skin_surf}
\end{equation}
It is worthwhile to notice that each of these contributions depends on  both WS radii and diffusivities
of neutron and proton distribution.

\section{Results}\label{sec:results}

\subsection{Performance of the ETF approximation on experimental data}

\begin{figure}
\begin{center}
\includegraphics[angle=0, width=0.89\columnwidth]{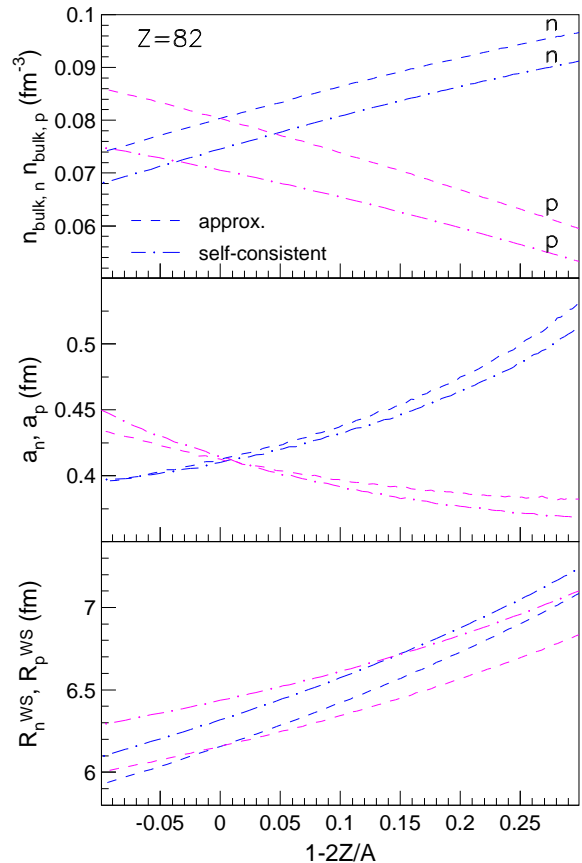}
\end{center}
\caption{ETF results corresponding to the ground state of Pb isotopes, for a representative  EDF model  (SLy4~\cite{SLY4}).
Variational parameters $n_{bulk,n}$, $n_{bulk,p}$ (top panel), 
$a_n$, $a_p$ (middle panel) and $R_n^{WS}$ and $R_p^{WS}$ (bottom panel), 
are plotted as a function of total isospin asymmetry.
The results obtained by considering the diffusivity of the charge distribution 
("self-consistent", Eq. (\ref{eq:exactCoulomb})) are confronted with those corresponding 
to the uniformly charged sphere approximation ("approx.").
}
\label{fig:ETF_variational}
\end{figure}

In order to visualize the overall performance of the ETF approximation, we consider in this section 
a single nuclear EDF model, namely the SLy4~\cite{SLY4} functional. 
We remind that the lot of data on which SLy4~\cite{SLY4} has been constrained 
includes binding energies and rms radii of doubly magic nuclei and
the equation of state of pure neutron matter of Ref. \cite{Wiringa_1988}. 
The last constraint guarantees a correct behavior at high isospin asymmetry.

In terms of average standard deviation on masses and radii, we obtain for the considered pool of spherical nuclei  
$\chi(B) = 4.9 $ MeV and $\chi(R_{\rm ch}) = 4.1 \cdot 10^{-2} $ fm.

\begin{figure}
\begin{center}
\includegraphics[angle=0, width=0.89\columnwidth]{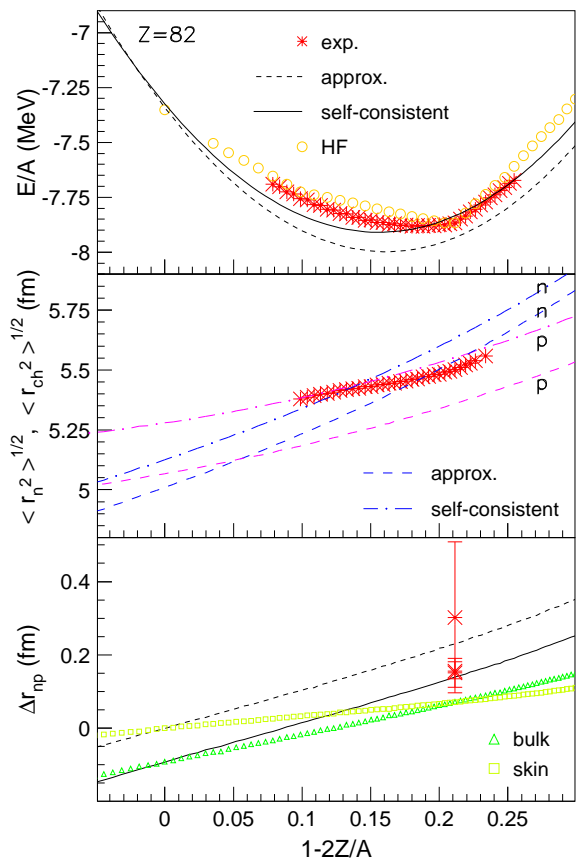}
\end{center}
\caption{ETF results corresponding to the ground state of Pb isotopes and SLy4~\cite{SLY4}.
Binding energy per nucleon (top panel), rms radii of neutron and charge distributions (middle panel)
and neutron skin thickness are plotted as a function of total isospin asymmetry.
When available, experimental masses \cite{AME2012} and charge radii \cite{Marinova} are plotted as well.
For neutron skin thicknesses of $^{208}{\rm Pb}$ the following experimental data are illustrated:
$0.1515 \pm 0.0197$ fm \cite{Trzcinska_PRL87},
$0.156^{+0.025}_{-0.021}$ fm \cite{Tamii_2011} and
$0.3012 \pm \left( 0.175 \right)_{\rm exp} \pm \left( 0.026 \right)_{\rm model}\pm \left( 0.005 \right)_{\rm strange}$
fm \cite{Abrahamyan_2012,Horowitz_2012}. 
As in Fig. \ref{fig:ETF_variational}, 
two methods for calculating the Coulomb energy are considered.
Neutron skin thickness decomposition into bulk and surface contributions according 
to eqs. (\ref{eq:skin_bulk}, \ref{eq:skin_surf})  is represented on the bottom panel for the
case in which the Coulomb energy is calculated self-consistently (open symbols).
}
\label{fig:ETF_observables}
\end{figure}

The results of total, i.e. nuclear plus electrostatic, energy minimization in the 
4-dimensional space $\{ n_{bulk, n}, n_{bulk,p}, a_n, a_p\}$
are plotted in Figs. \ref{fig:ETF_variational} and \ref{fig:ETF_observables}
for the isotopic chain of Pb as a function of the isospin asymmetry, $I=1-2Z/A$.
Two different methods are  used to calculate the Coulomb energy.
In one case it is calculated by accounting for the diffusivity of the proton distribution via 
eq. (\ref{eq:exactCoulomb}) ("self-consistent"). In the second, a uniformly charge distribution approximation 
is employed, which leads to $0.69 Z^2/A^{1/3}$ ("approx.").
The top and middle panels of Fig. \ref{fig:ETF_variational} present the evolution of each of the four
variational parameters as a function of $I$. The bottom panel presents the $I$-dependence
of the WS radii on neutron and proton distributions, obtained from particle number conservation. 
We can notice the important effect of a self-consistent treatment of Coulomb in the determination of the density profiles. 
In particular, the obtained bulk densities and diffuseness parameters are in good agreement with fits of 
HF density profiles with the same EDF \cite{Panagiota}, which comforts us on the quality of the approximation.

We can also see that WS radii of neutron and proton distributions have similar values, though strongly
dependent on $I$. This might suggest that the skin is mainly a surface effect for this calculation. 
However, this interpretation is not correct because the equivalent sharp radius 
$R_q^3=\left( 3 \int dr r^2 n_q(r) \right)/n_{bulk,q}$ 
is different from the 
WS radius parameter, $R_q^{WS}$, and effectively depends on the diffuseness of the profile \cite{Mondal_PRC93}. 
Moreover, as explicitly worked out in Ref. \cite{Francois2}, the diffuseness parameter itself depends 
in a highly non trivial way both on the gradient terms of the EDF and on the bulk properties of matter.

Fig. \ref{fig:ETF_observables} illustrates the total binding energy per nucleon (top panel),
rms radius of charge distribution (middle panel) and neutron skin thickness (lower panel) 
as a function of $I$. 
When available experimental data for binding energies \cite{AME2012} and charge radii \cite{Marinova}
are plotted as well.
For neutron skin thickness of $^{208}{\rm Pb}$ we display data from 
Refs. \cite{Trzcinska_PRL87,Tamii_2011,Abrahamyan_2012,Horowitz_2012}.
Self-consistent calculation of the Coulomb energy leads to a fair agreement with experimental data 
though a systematic over binding is obtained for nuclei with $I < 0.17$. 
Complete HF calculations from Ref. \cite{Debi}, performed in spherical symmetry, are also shown. 
Aside a residual deviation which can be ascribed to the choice of the functional and/or beyond mean-field effects,
HF calculation describe very well the experimental data.
Concerning the ETF calculations, we can see that missing higher $\hbar$ orders and the use of a parametrized 
density profiles lead to a deviation with respect to the experimental data which 
is larger then the one of the HF calculation.
The energy error is however very small for $^{208}{\rm Pb}$ and neighboring nuclei. 
This justifies the method described in IIB and employed to build
meta-modelling EDF based on best fit of properties of spherical nuclei.

The performances of the ETF approximation when Coulomb is consistently included in the variation, 
can be judged also from the agreement of rms radii of charge distributions with experimental data.
As one may see in the middle panel of Fig. \ref{fig:ETF_observables}, the overall accord is good.
The most important deviations, of the order of 0.05 fm, are obtained for $I>0.17$. 
This deviation is comparable to the one obtained with complete ETF or DFT calculations 
in the absence of deformation~\cite{Buchinger,Patyk}, and can be ascribed to the choice of the 
functional and/or to beyond mean field effects.
Neutron skin thickness presents a linear dependence on $I$ irrespective how Coulomb was calculated.
As easy to anticipate, the consistent displacement of neutron and proton distributions, 
due to the Coulomb repulsion, leads to values of the neutron skin thickness lower 
than those obtained in the simplifying approximation.
It is interesting to remark that while the Coulomb effect decreases the neutron skin, the different diffuseness 
of the proton and neutron density profiles tends to increase it. 
As a consequence, the two effects partially cancel and the global result is close to our 
previous calculations \cite{Debi}, 
where both effects were neglected in order to obtain analytic approximations.
The bottom panel depicts also the bulk and surface contribution to the skin thickness \cite{Mondal_PRC93}, 
calculated according to eqs. (\ref{eq:skin_bulk}, \ref{eq:skin_surf}). 
One notices that, for $^{208}{\rm Pb}$, they contribute equally to the total thickness while in neutron-richer 
(neutron-poorer) isotopes it is the bulk (surface) term that dominates.
Given the relatively low $L_{sym}$=46 MeV value of SLy4 \cite{SLY4},
this result is in good agreement with the droplet model (DM) calculations of Ref. \cite{Centelles_PRC82}, 
where the dominance of bulk/surface contributions was shown to be linked to the value of  $L_{sym}$.

\subsection{Correlations between nuclear observables and parameters of nuclear matter}

\begin{figure}
\begin{center}
\includegraphics[angle=0, width=0.99\columnwidth]{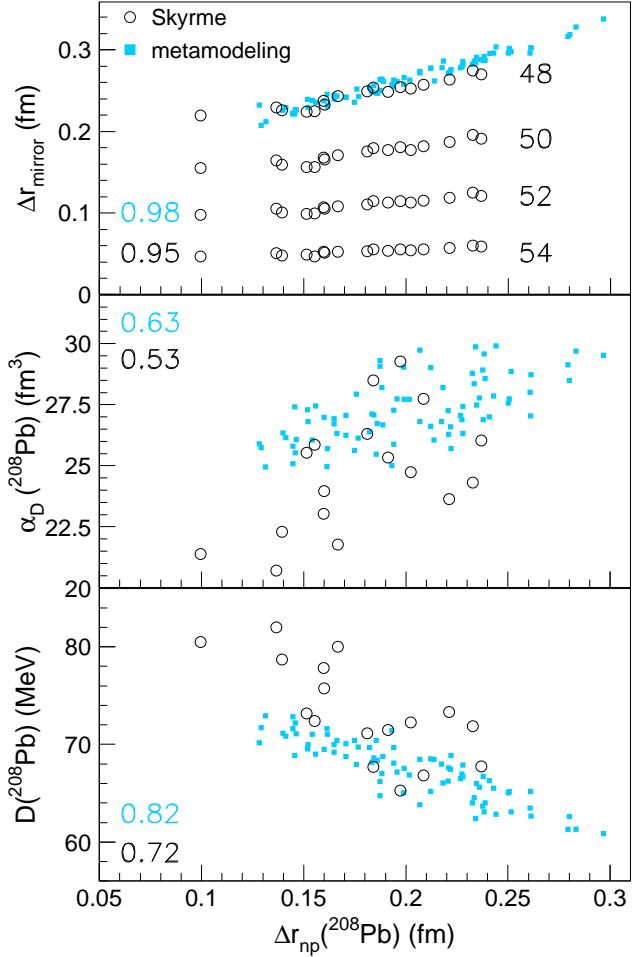}
\end{center}
\caption{Correlations between neutron skin thickness in $^{208} {\rm Pb}$ and
differences in the proton radii of mirror nuclei 
$R_p (^{48}{\rm Ni})-R_p (^{48}{\rm Ca})$ (top),
electric dipole polarizability of $^{208}{\rm Pb}$ (middle) and
IVGDR energy constant of $^{208} {\rm Pb}$ (bottom).
Results corresponding to Skyrme and meta-modelling are represented with
open circles and, respectively, solid squares.
Skyrme predictions corresponding to differences in the proton radii of mirror nuclei
$R_p (^{50}{\rm Ni})-R_p (^{50}{\rm Ti})$, 
$R_p (^{52}{\rm Ni})-R_p (^{52}{\rm Cr})$,
$R_p (^{54}{\rm Ni})-R_p (^{54}{\rm Fe})$ are also plotted in the top panel. 
Numbers on l.h.s. of each plot correspond to Pearson correlation coefficients between the observables
plotted on the axis. Upper (lower) values: meta-modelling (Skyrme). 
}
\label{fig:skin-otherobs}
\end{figure}

The correlation between the neutron skin thickness of $^{208}{\rm Pb}$ and $L_{sym}$ 
has been reported in the past years in many different studies based on density functionals 
\cite{furnstahl,rocamaza12,Mondal_PRC93}, semi-classical approaches \cite{Warda09,Centelles_PRC82}, 
as well as DM \cite{Centelles_PRC82}.
  
More recently, the existence of other correlations with various isovector modes of 
collective excitation was suggested, 
namely electric dipole polarizability \cite{Tamii_2011,rossi,birkhan}, 
isovector giant dipole resonance (IVGDR) \cite{VanIsacker_1992},
isovector giant quadrupole resonance (IVGQR) \cite{henshaw},
pygmy dipole resonance (PDR) \cite{VanIsacker_1992,Baran_2013,carbone}, 
anti-analog giant dipole resonance (AGDR) \cite{Krasznahorkay_ActaPolonica2013,Krasznahorkay_2013,Cao_2015}.
A correct description of these modes demands a dynamical treatment in the framework of 
linear response theory and is beyond the purpose of this work. 
However, simplified expressions were proposed.
An example in this sense is given by Ref. \cite{RocaMaza_2013} which relates
the electric dipole polarizability of a nucleus of mass number $A$ and 
isospin asymmetry $I$,
\begin{equation}
\alpha_D=\frac{\pi e^2}{54} \frac{A \langle r^2 \rangle}{E_{sym}} \left( 1+\frac53 \frac{E_{sym}-a_{sym}}{E_{sym}}\right). 
\label{eq:alphaD}
\end{equation}
with the ground state symmetry energy in the local density approximation \cite{Mondal_PRC93}
\begin{equation}
a_{sym}(A)=\frac{4\pi}{A I^2}\int_0^\infty dr \, r^2 n(r)\delta^2(r)e_{sym}(n(r)),
\label{eq:asym}
\end{equation}
where $e_{sym}=(1/2)\partial^2 e(n,\delta)/\partial\delta^2|_{\delta=0}$
represents the local symmetry energy.
Another example is offered by Ref. \cite{Blocki_2013} which expresses
the IVGDR energy constant in terms of symmetry energy, saturation density 
and surface stiffness coefficient, $Q_{stiff}$, as,
\begin{equation}
D=D_{\infty}/\sqrt{1+3 E_{sym} A^{-1/3}/Q_{stiff}}, \label{eq:D}
\end{equation}
where
$D_{\infty}=\sqrt{8 \hbar^2 E_{sym}/\left( m r_0^2\right)}$ and $r_0^3=3/\left( 4 \pi n_{sat}\right)$.
The surface stiffness coefficient measures the resistance of the asymmetric semi-infinite nuclear matter
against separation of neutrons and protons to form a skin and is typically performed within HF or ETF approaches. 
Such calculations showed some sensitivity of $Q_{stiff}$ to the calculation procedure
\cite{Kolehmainen_NPA_1985,Centelles_NPA_1998}
as well as significant correlations with the symmetry energy and its first and second order derivatives
\cite{Treiner_AnnPhys_1986,Warda09}.
Different approximation formulas have been proposed.
Some of them express $Q_{stiff}$ in terms of a number of nuclear matter parameters and are based on fits of
HF or ETF calculations performed using different EDFs.
Within the Liquid Drop Model, Ref. \cite{Brack_1985} calculates $Q_{stiff}$ from calculations of 
finite nuclei disregarding the Coulomb interaction.
In the present work we adopt the expression,
$Q_{stiff}=9 E_{sym} A^{-1/3}/4/ \left( E_{sym}/a_{asym}-1\right)$,
obtained by equating the ground state symmetry energy given by eq.~(\ref{eq:asym}) 
with the corresponding DM expression \cite{Mondal_PRC93}.
For the case of $^{208}{\rm Pb}$ its accuracy is of the order of 10\%, which leads to a relative error
of 2\% on the IVGDR energy constant of $^{208}{\rm Pb}$ calculated according to eq.~(\ref{eq:D}).
This small uncertainty only marginally affects the correlation between the macroscopically derived 
IVGDR energy constant and various properties associated with the finite nuclei or the nuclear matter.
However, more important distortions might come from the nature of the approximation itself, namely
the use of macroscopic expressions in case of dynamical quantities. 
Such distortions apply to both $\alpha_D$ and $D$.

Another interesting observable, potentially linked to the isovector EoS parameters, is given by the difference 
between the proton radii $R_p=\langle r_p^2 \rangle^{1/2}$ of mirror nuclei \cite{Brown_2017,Piekarewicz_2017}. 
This observable has the interesting feature of being directly accessible from a variational calculation 
without any extra model assumption. Moreover, it is much more accessible experimentally than the neutron skin, 
which demands the measurement of the neutron distribution.
 
The correlation between the proton radii differences in mirror nuclei and
electric dipole polarizability on one hand and neutron skin thicknesses on the other hand has
been addressed in Refs. \cite{Brown_2017,Piekarewicz_2017,RocaMaza_JPG_2015,RocaMaza_PRC_2015}.
Ref. \cite{Blocki_2013} focused on the nuclear symmetry energy dependence of the IVGDR energies by considering
a series of Skyrme interaction potentials.
The correlations between neutron skin thickness, electric dipole polarizability and IVGDR energy constant
of $^{208}{\rm Pb}$
and proton radii difference for $A=48$ mirror nuclei are investigated in
Fig. \ref{fig:skin-otherobs} for both meta-modelling EDF and Skyrme functional.
For completeness, Skyrme predictions corresponding to differences in the proton radii of
$A=50, 52, 54$ and $\Delta r_{np}(^{208}{\rm Pb})$ are also plotted in the top panel.
In the case $\Delta r_{\rm mirror}$ vs. $\Delta r_{np} (^{208}{\rm Pb})$. 
meta-modelling EDF lead to a strong correlation, with a Pearson correlation coefficient of 0.98.
A moderate correlation is obtained for $D(^{208}{\rm Pb})$ vs. $\Delta r_{np} (^{208}{\rm Pb})$.
A poor correlation is found between the dipole polarizability and neutron skin thickness. 
Skyrme functionals provide very similar results.
Very strong correlations are obtained only between $\Delta r_{np}(^{208}{\rm Pb})$ and 
proton radii differences in mirror nuclei with $A=48, 50, 52, 54$. 
This result is in agreement with Refs. \cite{Brown_2017,Piekarewicz_2017}.
The correlation between electric dipole polarizability and $\Delta r_{np}(^{208}{\rm Pb})$ is loose,
in agreement with Ref. \cite{RocaMaza_PRC_2015}. Ref. \cite{RocaMaza_PRC_2015} has actually evidenced
that a much better correlation holds between $\Delta r_{np}$ and $\left( \alpha_D E_{sym} \right)$,
as expected from eq. (\ref{eq:alphaD}).
Finally Skyrme functionals lead to medium strength correlations between IVGDR energy constant 
and neutron skin thickness of $^{208}{\rm Pb}$. This result can be understood considering 
the $L_{sym}$- and $K_{sym}$- dependence of the $D$ quantity via $Q_{stiff}$.

\begin{figure}
\begin{center}
\includegraphics[angle=0, width=0.99\columnwidth]{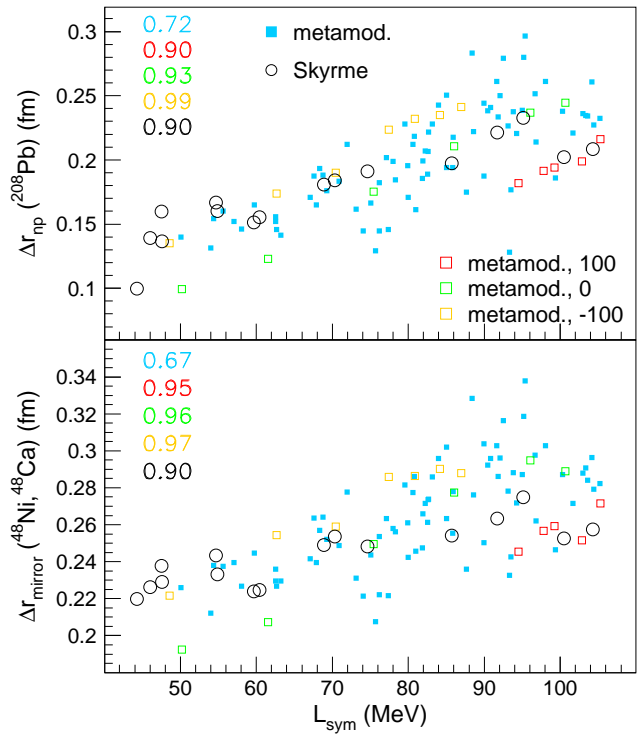}
\end{center}
\caption{Correlations between $L_{sym}$ and $\Delta r_{np} (^{208}{\rm Pb})$ (top panel)
and $L_{sym}$ and $R_p (^{48}{\rm Ni})-R_p (^{48}{\rm Ca})$ (bottom panel).
In addition to meta-modelling EDF plotted in the previous figures here we consider also 
meta-modelling EDF with fixed values of $K_{sym}=100, 0, -100$ MeV.
The numbers on the l.h.s mention the Pearson correlation coefficients in the following order:
meta-modelling with freely varying $K_{sym}$,
meta-modelling with $K_{sym}=100, 0, -100$ MeV and Skyrme.
}
\label{fig:L_goodcorrel}
\end{figure}

We now turn to test the sensitivity of the observables to the different isovector parameters of the EoS. 
In a previous work \cite{Debi}, a full Bayesian analysis of the correlation matrix was performed, 
though with a more simplified version of the ETF meta-modelling, which did not include the self-consistent 
treatment of Coulomb nor the definition of $(n_{bulk,n}, n_{bulk,p} ,a_n, a_p)$ as independent variational variables. 
In that study, it was shown that the neutron skin is only sensitive to the $L_{sym}$ parameter. 
The present calculations, with a more sophisticated treatment of the ETF meta-modelling, 
confirm the results of our previous work.

The correlation between the neutron skin in $^{208}{\rm Pb}$ and the $L_{sym}$ parameter
is shown in the top panel of Fig. \ref{fig:L_goodcorrel}. The lower value of the correlation coefficient with respect 
to the results of Ref. \cite{Debi} can be understood from the fact that the difference between 
the neutron and proton diffusivity was neglected in Ref. \cite{Debi}. This value is also lower than the one 
corresponding to Skyrme functionals, as well as to the ones reported by most analyses in the literature using  
specific energy functionals \cite{RocaMaza_JPG_2015,Mondal_PRC93,Warda09,Centelles_PRC82,Reinhard16}.
The higher dispersion of the meta-modelling is due to the fact that the different EoS parameters are fully independent 
in the meta-modelling approach. As already observed in Ref. \cite{Debi}, though the EoS parameters 
are all influential in the calculation of nuclear masses and radii, the constraint on those quantities 
does not generate correlations among the EoS parameters because compensations can freely occur.

To demonstrate this statement, we have generated models with arbitrary fixed values of $K_{sym}$ 
fulfilling the same criteria imposed to the global set of models, see Section IIB.
The resulting correlations are shown in Fig. \ref{fig:L_goodcorrel} for three cases $K_{sym}=-100,0,100$ MeV. 
We can observe that the correlation between $^{208}{\rm Pb}$ and $L_{sym}$ is greatly improved when $K_{sym}$ is fixed. 
In the case of Skyrme functionals, $K_{sym}$ can largely vary but its value is positively correlated to 
$L_{sym}$ because of the specific function form of the density dependent term in Skyrme interactions 
(see Figure \ref{fig:EoSParamCorr} (a)).
As a consequence, the Skyrme results interpolate the more general meta-model ones and the
correlation coefficient is only slightly less than those corresponding to meta-model EDF with fixed 
$K_{sym}$-values.

The bottom panel of Fig. \ref{fig:L_goodcorrel} summarizes the analyses done above but for the correlation
between the proton radii difference in $A=48$ mirror nuclei and $L_{sym}$. The conclusions are similar:
strong (poor) correlations exist in the case of Skyrme functionals and meta-modelling EDF with fixed
$K_{sym}$-values (meta-modelling EDF with freely varying $K_{sym}$). 

\begin{figure}
\begin{center}
\includegraphics[angle=0, width=0.99\columnwidth]{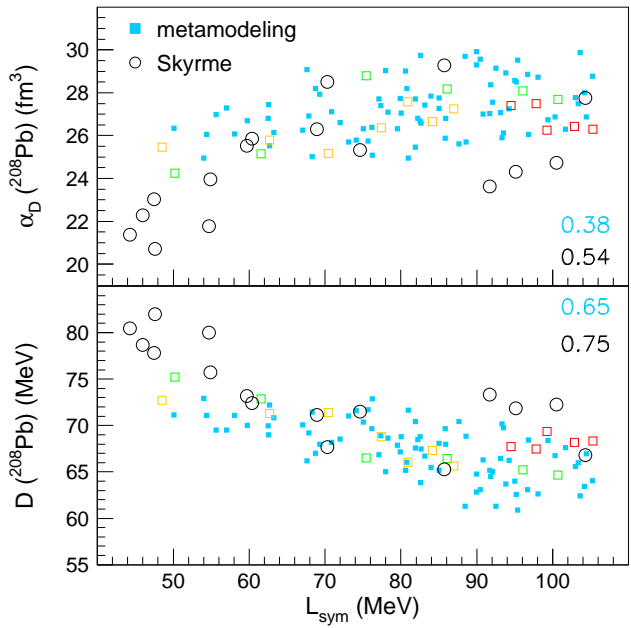}
\end{center}
\caption{Correlations between $L_{sym}$ and $\alpha_D (^{208}{\rm Pb})$ (top) and
$L_{sym}$ and $D(^{208}{\rm Pb})$ (bottom).
Numbers on r.h.s. of each plot correspond to Pearson correlation coefficients between the observables
plotted on the axis. Upper (lower) values: meta-modelling (Skyrme). 
As in Fig. \ref{fig:L_goodcorrel} meta-modelling EDF with fixed
values of $K_{sym} = 100, 0, -100$ MeV are plotted as well.
}
\label{fig:L_poorcorrel}
\end{figure}

The correlations of the dipole polarizability and IVGDR energy constant of $^{208}{\rm Pb}$
with $L_{sym}$ are reported in Figure \ref{fig:L_poorcorrel}, 
for the meta-modelling and for the selected Skyrme functionals. 
As in Fig. \ref{fig:L_goodcorrel} meta-modelling EoS with fixed values of $K_{sym}=-100,0, 100$ MeV
are also considered.
As one may see, $\alpha_D (^{208}{\rm Pb})$ and $D(^{208}{\rm Pb})$ show less correlation with
$L_{sym}$ than with $\Delta r_{np}(^{208}{\rm Pb})$, when meta-modelling EDF are employed.
At variance with this, Skyrme functionals provide for $\alpha_D (^{208}{\rm Pb})$ and $D(^{208}{\rm Pb})$
almost the same degree of correlation with $L_{sym}$ as with the neutron skin of $^{208}{\rm Pb}$.
A word of caution is nevertheless in order.
The accurate calculation of these two dynamical quantities is possible only within the linear response theory.
The Eqs. (\ref{eq:alphaD}, \ref{eq:D}) presently employed rely on approximations and are, thus, expected to
distort the sensitivity to nuclear matter EoS.

In Ref. \cite{Debi} the isoscalar and isovector parameters of the meta-modelling EDF have been determined by fits of 
experimental binding energies of symmetric nuclei with masses $20 \leq A \leq 100$ and full isotopic chains of 
Ca, Ni, Sn and Pb. We have tested that the conclusions drawn above and the degrees of correlation remain the same
if the pool of nuclei on which the parameters of the EDF are determined is replaced by the one 
considered in Ref. \cite{Debi}.

\section{Conclusions}\label{sec:conclusions}

In this paper, we have explored the influence of the different isovector empirical EoS parameters on 
some properties of atomic nuclei, namely 
neutron skin thickness, difference in proton radii of mirror nuclei,
dipole polarizability and the IVGDR energy constant of $^{208}{\rm Pb}$.

The analysis was done within a recently proposed meta-modelling technique \cite{JM1,Debi}. 
Varying the parameters of the meta-modelling, it is possible to reproduce existing relativistic 
and non-relativistic EDF, as well as to consider novel density dependencies which are not explored 
by existing functionals.  
With respect to our previous work Ref. \cite{Debi}, we have improved the ETF formalism {employed to extract 
nuclear masses and radii out of a given EDF: the Coulomb interaction is consistently included 
in the variational procedure, and the bulk densities and diffuseness parameters of the density profiles 
are treated as independent variational parameters. 
These improvements allow a better description of nuclear radii and the nuclear skin. The correlation between this 
latter observable and the slope of the symmetry energy $L_{sym}$, already reported in numerous 
studies in the literature with different EDFs as well as many body techniques, is confirmed by our study.

However, we show that the quality of this correlation is considerably worsened if we allow independent 
variations of the curvature parameter $K_{sym}$ with respect to the slope $L_{sym}$, while this was not observed 
in previous studies probably because in most existing functionals such correlation exists. 
We conclude that it will be very important to constrain the curvature parameter with dedicated studies, 
in order to reduce the confidence intervals of EoS parameter and allow more reliable extrapolations 
to the higher density domain.

We have shown that the condition of a reasonable reproduction of nuclear masses and radii does not necessarily 
imply any strong correlation between $L_{sym}$ and $K_{sym}$. For this reason, it is possible that the existing 
correlation in the Skyrme EDF might be spuriously induced 
by the arbitrariness of the functional form, particularly the density dependent term.
However, as suggested in Ref. \cite{Mondal17,Tews17}, such a correlation might also be physical and linked to the 
fact that Skyrme EDF are derived from a pseudo-potential which satisfies some basic physical properties, 
which is not the case of the more general meta-modelling. 
To answer this question, it will be important to evaluate and possibly constrain this correlation on 
ab-initio calculations of neutron matter \cite{Tews17}. This work is presently in progress.

\end{document}